\documentclass[conference]{IEEEtran}
\usepackage{graphicx}
\usepackage{soul} 
\usepackage{url}
\usepackage{xcolor}
\usepackage{booktabs}
\usepackage{caption}
\usepackage{subcaption}
\usepackage{amsmath}
\usepackage{multirow}
\usepackage{rotating}
\usepackage[T1]{fontenc}
\usepackage[normalem]{ulem}
\usepackage{siunitx}
\usepackage[color=blue!10,textsize=footnotesize,textwidth=25mm]{todonotes}
\usepackage{hyperref}

\hypersetup{
  colorlinks   = true,    
  urlcolor     = blue,    
  linkcolor    = blue,
  citecolor    = blue  
}

\usepackage{soul, color}
\usepackage{color}

\newcounter{JGHCommentsCounter}

\newcounter{BPUCommentsCounter}

\newcounter{EHZCommentsCounter}

\begin{document}
\title{Towards Measuring the Impact of Technical Debt on Lead Time: An Industrial Case Study}

\author{
    \IEEEauthorblockN{Bhuwan Paudel, Javier Gonzalez-Huerta, Ehsan Zabardast, Eriks Klotins}
    \IEEEauthorblockA{Software Engineering Research Lab SERL,\\Blekinge Institute of Technology, Karlskrona, Sweden\\
    \{bhuwan.paudel, javier.gonzalez.huerta, ehsan.zabardast, eriks.klotins\}@bth.se}
}
\maketitle

\begin{abstract}

Background: Software companies must balance fast value delivery with quality, a trade-off that can introduce technical debt and potentially waste developers' time. As software systems evolve, technical debt tends to increase. However, estimating its impact on lead time still requires more empirical and experimental evidence.

Objective: We conduct an empirical study investigating whether technical debt impacts lead time in resolving Jira issues. Furthermore, our aim is to measure the extent to which variance in lead time is explainable by the technical debt.

Method: We conducted an industrial case study to examine the relationship in six components, each of which was analyzed individually. Technical debt was measured using SonarQube and normalized with the component's size, while lead time to resolve Jira issues was collected directly from Jira.

Results: We found a set of mixed results. Technical debt had a moderate positive impact on lead time in two components, while we did not see a meaningful impact on two others. A moderate negative impact was found in the remaining two components.

Conclusion: The findings show that technical debt alone can not explain all the variance in lead time, which ranges from 5\% up to 41\% across components. So, there should be some other variables (e.g., size of the changes made, complexity, number of teams involved, component ownership) impacting lead time, or it might have a residual effect that might manifest later on. Further investigation into those confounding variables is essential. 

\end{abstract}


\section{Introduction}
The relentless demand for new features forces software development organizations to rethink their development practices. The delivery of high-quality software and support, as well as the fast value to the customer, should be one of the core priorities of software organizations. However, this can only be achieved if the software quality is maintained, significantly impacting the overall software development lifecycle~\cite{besker2017time}. 

Software companies usually invest their time and effort into refactoring and improvement activities to identify and remove technical issues that could potentially improve software quality~\cite{lenarduzzi_technical_2021}, although these refactoring activities do not always succeed in removing technical debt items, but rather might introduce new ones~\cite{zabardast_refactoring_2020}. Different studies (e.g.,~\cite{besker2019software, tornhill_code_2022,8094405}) have reported that software development time is often wasted due to technical debt ranging from 23\% up to 42\%. So, technical debt may directly affect software development by reducing development speed and increasing maintenance costs now or in the future~\cite{10449672}.

In the pursuit of faster time-to-market and rapid delivery of new features and products, software companies have often deprioritized maintaining code quality~\cite{tornhill_code_2022,giardino_software_2015}, preferring release speed. The consequences of this trend from the software quality perspective have yet to be explored~\cite{6224279}. According to a study by Lehman~\cite{lehman1980programs}, software systems must keep evolving and adapting to the environment. However, doing so can increase complexity and decrease quality if not appropriately maintained~\cite{yu2013empirical}. In addition, some empirical evidence (e.g.,~\cite{9234106, digkas2017evolution}) mentions that technical debt tends to increase over time due to the increased size and/or complexity with reduced quality of software systems. Consequently, managing technical debt is becoming a major task that plays a significant role in maintaining software quality~\cite{avgeriou_reducing_2015}. When code quality declines, the time to work on new features gets shortened, and the time required to implement them is extended because the code will be more complicated to change~\cite{tom_exploration_2013}.

 Manual approaches become unrealistic and painful as complexity and size grow more and more~\cite{10.1007/978-3-031-04115-0_1}. 
 Data-driven approaches are promising, but we still need more evidence. A few studies (e.g.,~\cite{lenarduzzi_technical_2021, tornhill_code_2022}) have initiated a data-driven approach to investigate the impact of code quality and technical debt on the lead time. However, further exploration in other contexts, such as at different levels of abstraction and industrial evidence, is needed. In this paper, we take a step in providing such evidence by investigating the impact of technical debt on lead time. To the best of our knowledge, this is the first case study on a proprietary, industrial setting on this particular topic, thus demonstrating industrial relevance while being aligned with the interests of the research community and practitioners. 

In this study, we define the lead time as the sum of time developers spend in different statuses associated with resolving tickets or developing new features (i.e., aggregated sum of time in progress, code review, and testing). This approach to measuring lead time differs from previous studies (e.g.,~\cite{lenarduzzi_technical_2021, tornhill_code_2022}), and we believe this helps to obtain the actual time developers spend resolving Jira issues more accurately because technical debt can also hinder time spent during the code review and testing. Additionally, the technical debt measure was from SonarQube, normalized by the component's size to get technical debt density (TDD). TDD is a metric that has been employed in several other studies (e.g.,~\cite{9234106, 10011486, al2019evolution}).

Our results suggest that although technical debt, in theory, can be a factor that might have a certain impact on lead time to some extent in certain components, it could not entirely explain the variance in lead time across all components studied. Moreover, our study raises concerns that other confounding variables, such as the size of the changes made, complexity, number of teams involved, and component ownership, might be at play. These findings indicate the need for further investigation with more fine-grained analyses.


The rest of the paper is organized as follows. Section~\ref{sec:background} provides the background for technical debt and lead time. Section~\ref{sec:relatedwork} presents the related work. Section~\ref{sec:methodology} outlines the study design, from the research question to data analysis. Section~\ref{sec:results} reflects the results of the study, and Section~\ref{sec:discussion} provides a discussion. Section~\ref{sec:threats} identifies threats to the validity of our research, and Section~\ref{sec:conclusions} concludes the study.

\section{Background}\label{sec:background}
\subsection{ Technical Debt (TD)}

Technical debt is a metaphor widely used to reason about the consequences of poor software development practices and design decisions~\cite{tom_exploration_2013,ernst_measure_2015}. Cunningham~\cite{cunningham_wycash_1992} initially introduced the concept of technical debt to discuss the potential long-term negative consequences of sub-optimal solutions adopted during software development. TD will eventually have to be repaid but generate interest in the long run,~\cite{zabardast_refactoring_2020, besker_systematic_2016}. 
TD was initially the most used term in economics, but it has now been widely adopted in software maintenance and evolution~\cite{lenarduzzi_technical_2021}.

The survey results from the study~\cite{jaspan2023defining} found that no single metric effectively identifies the leading indicators of TD~\cite{10449672}. Teams must select and adjust metrics based on context, as the existence of TD is context-dependent. Lenarduzzi et al.~\cite{lenarduzzi_technical_2021} mentioned that estimating TD with a precise value is difficult due to its complex phenomenon. However, it is possible to estimate several of its properties. 

The use of static code analysis tools in the industry and academia for assessing code quality and understanding TD has been gaining traction~\cite{ernst_measure_2015, lenarduzzi_survey_2020}. SonarQube\footnote{\url{https://www.sonarsource.com/products/sonarqube/}} is one of the most used open-source static code analysis tools for measuring software code quality~\cite{lenarduzzi_technical_2019}. It is an automatic code review tool that systematically scans the source code to ensure it adheres to a specific set of coding rules, and its violations are reported as technical debt issues, using different metrics such as lines of code or code complexity~\cite{baldassarre_diffuseness_2020}.



SonarQube generates ``issues'' when it detects violations of coding rules or quality gates violations. These issues are categorized as \emph{code smells}, \emph{bugs}, or \emph{vulnerabilities}.
Code smells symbolize issues related to maintainability, while bugs and vulnerabilities indicate reliability and security concerns, respectively~\cite{baldassarre_diffuseness_2020}. It also classifies these issues based on their severity score from lowest to highest, such as info, minor, major, critical, and blocker. SonarQube estimates the time required to resolve each TD item (sonar issues), which is defined as the remediation time for that particular item. 

We collected the Sonar remediation time as a metric for TD for the past two years (2022-2023). The TD per component is calculated by adding up the TD present in each file for the respective component. We normalized the amount of TD per component with the corresponding component's size, called the technical debt density (TDD) as the independent variable for this study. The concept of TDD has been used in many previous studies, such as~\cite{9234106, 10011486, al2019evolution}. TDD enables us to reason about the evolution and the impact external factors might have on the accumulation of TD~\cite{10011486}.

\subsection{Lead Time}\label{section:background_leadtime}

Jira~\footnote{\url{https://www.atlassian.com/software/jira}} is an issue-tracking and agile project management tool for software development teams developed by Atlassian. Jira issues include feature development and other software maintenance activities. These other software maintenance activities might consist of the resolution of Jira issues such as bugs, code smells, or vulnerabilities that SonarQube identified. However, as opposed to~\cite{lenarduzzi_technical_2021}, which focuses only on the resolution of technical debt items (TDIs), we measure lead time for every Jira issue. It is important to note that Jira provides two types of time during issue resolution~\cite{tornhill_code_2022}: First, cycle time (time from the beginning to the end of a specific action, i.e., time when issues are marked as ``In progress''). Second, lead time, which covers the entire duration from receiving a request until its completion, including any waiting time.

\begin{figure}[htbp]
    \centering
    \includegraphics[width=1\linewidth]{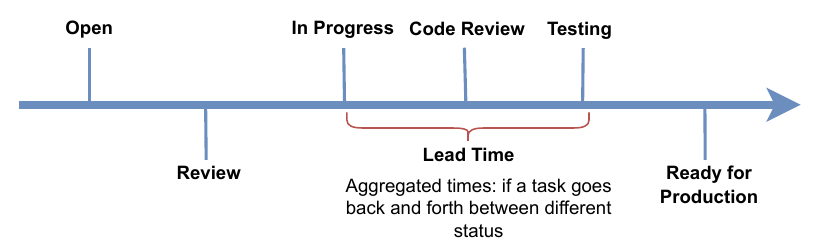}
    \caption{Lead time calculation for Jira Issues.}
    \label{lead time}
\end{figure}

However, we based our lead time measurement on how the company tracks lead time. As shown in Figure~\ref{lead time}, we calculated lead time by aggregating the sum of time spent in different statuses associated with resolving Jira issues or developing new features.
 
The lead time includes: \textbf{In Progress:} Total time the task (i.e., Jira issue) is in development. \textbf{In Code Review:} Total time the task is in the code review process, as reported in Jira. \textbf{In Testing:} Total time the task is under test after it has gone through the code review process. We gathered all Jira issues and lead time data for each component. We then created a map of the issues and their corresponding lead time data using \emph{issueKey}. This helped us get the lead time data for all Jira issues for specific components.

\section {Related Work}\label{sec:relatedwork}

Besker et al.~\cite{besker2017time} found a linear relationship between quality issues and the amount of time wasted. It means that every time developers face quality issues, more software development time is wasted. As mentioned in Tornhill \& Borg~\cite{tornhill_code_2022}, up to 42\% of developers' time is spent managing TD, where most expenses are incurred due to the low-quality code costing software companies around \$85 billion annually~\cite{noauthor_developer_2018}. 

Participants in a study by Tom et al.~\cite{tom_exploration_2013} reported that one of the first impacts of TD noticed is a slowdown in development speed. Liu et al.~\cite{5680918} suggest that the refactoring effort can be reduced by 17.64 to 20\% when bad smells are detected and resolved in time. In addition, resolving Jira issues in code in the alert category takes longer time than in the healthy code, potentially doubling the time to market~\cite{tornhill_code_2022}. 
In contrast, TD sometimes can be used strategically by trading off quality with productivity, such as speeding up the development of new features, yielding the benefit of faster time to market~\cite{10.1145/1985362.1985370, li_systematic_2015}, and increasing the business value. However, if the interest rate of TD is too high, this strategy may become less effective or even detrimental. Therefore, software managers need to carefully weigh the potential benefits and interests of incurring TD when it comes to planning software projects~\cite{10.1145/1985362.1985370}.

Tornhill \& Borg~\cite{tornhill_code_2022} investigated the relationship between code quality on a file level and the cycle time in development. They found that code quality significantly impacts development time. According to their findings, developers spend relatively more time resolving issues in low-quality source code. Their result shows that development time for Jira issues in the alert category code is, on average, 124\% more than in the healthy category code and 78\% more than in the warning category code. Poor source code quality can increase uncertainty in the time required to resolve issues. Conversely, high-quality source code can reduce this uncertainty~\cite{tornhill_code_2022, borg_u_2023}.

Similarly, Lenarduzzi et al.~\cite{lenarduzzi_technical_2021} initiated a data-driven approach to estimate the TD and its impacts on the variation in the lead time for resolving Jira issues with the aim of improving the estimation and prioritization of removing TD, based on its impact. However, study~\cite{lenarduzzi_technical_2021} did not find a meaningful correlation between lead time and violations. 

Our study addresses the gaps in previous studies. Unlike the study by Tornhill \& Borg~\cite{tornhill_code_2022}, we do not rely solely on `time-in-development' as we think TD can also hinder time spent during the code review and testing process. Lenarduzzi et al.~\cite{lenarduzzi_technical_2021} focuses only on the resolution of technical debt items while we analyzed lead time for all Jira issues. In~\cite{lenarduzzi_technical_2021}, authors calculated lead time simply from the inducing and fixing commits’ timestamps reported by SonarQube. First, this approach might be inaccurate since the fixing commit reported by SonarQube might be the merge commit squashing several other commits, and it is hard to distinguish whether this is the actual lead time to fix the TD item. The method we used to calculate lead time, particularly the time developers spent in three different statuses, as mentioned in Section~\ref{section:background_leadtime}, provides a more accurate measurement technique for lead time and can yield better and more empirical evidence.

Finally, we conducted separate analyses for each component. This approach eliminates the influences of some other variables, such as the size and complexity of changes made, the component's owner, and other contextual factors that could arise while putting all components together. 


\section{Methodology}\label{sec:methodology}
\subsection{Study Design}
This study considers each component as an individual case and analyzes them separately. Figure~\ref{fig: Study design} below illustrates our empirical study design based on the case study guidelines defined by~\cite{runeson_guidelines_2009}.

\begin{figure}[htbp]
    \centering
    \includegraphics[width=1\linewidth]{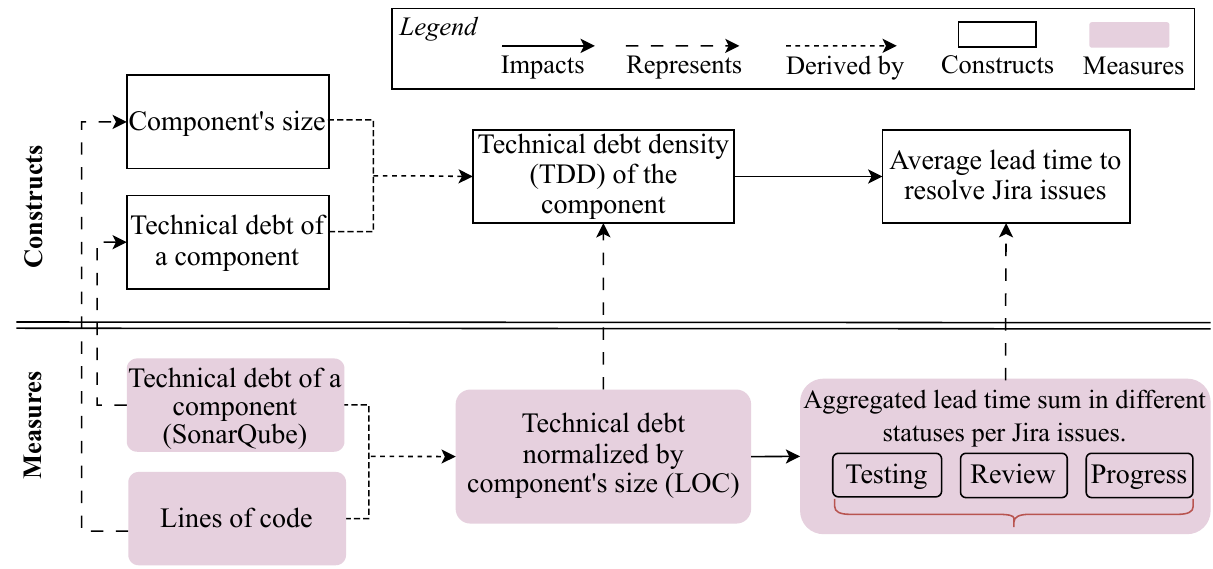}
    \caption{Illustration of the study design.}
    \label{fig: Study design}
\end{figure}

Our goal is to investigate whether a relationship exists between the TD in a specific component, as measured by a static analysis tool, normalized by the component's size, i.e., technical debt density (TDD), and the time taken to resolve Jira issues associated with that component. Our primary focus is to derive the relationship between the amount of TD density in a component and the average lead time to fix a Jira issue in that particular component. Based on this goal, this study examines the following research question: \textbf{RQ: \textit{How does technical debt impact the lead time?}}\label{sec:RQ}

We hypothesize that the TD present in a particular component, to some extent, impacts the average lead time required to resolve Jira issues in that component. If we can establish a relationship between these two variables, we should be able to determine to what extent the presence of TDD affects the lead time for resolving Jira issues or how much of the variation in lead time can be explained by TDD. Therefore, we first examine the relationship between these two variables and then assess the strength of the correlation by calculating the R-squared value. This value explains the percentage of variation in lead time that TDD can explain.

\subsection{Context}

We conducted this study with one of our industry collaborators, and the collected datasets are from SonarQube and Jira. The case has been selected by convenience. The company is a large and well-matured software development company that mostly develops products based on financial and accountancy services (i.e., the Fintech area). Java is the main programming language. Initially, we started with ten components and later excluded those with less than twenty months of active development in the last two years (2022-2023). The details of selected components for this study after exclusion are presented in Table~\ref{components}. 


\subsection{Data Collection}

We collected data from two primary sources: SonarQube and Jira. TD data was collected from the Company's SonarQube endpoints at the beginning of each month. The dataset includes information about violated rules, status, issue types, severity, size, complexity, number of violations, effort (TD), and the files and components involved. Additionally, this data contains the timestamp of when the issue was created, the updated date, and the fixed timestamp if it has been resolved. 

We only included issues with open status, as we intend to calculate the total TD present in a specific month and component. We then added the effort (TD) of all issues throughout the month in that component to get the total TD value (minutes) for components over a specific month. TD was normalized by the size of the corresponding components and multiplied by a thousand to get TDD per thousands of lines of code (KLOC), which is an independent variable in this study. 

Although it would be possible to assess and perform the analysis on individual Jira issues and the TD at that point in time, given the variability of individual data points, we opted for a higher-level aggregated view to capture general trends. 
Many other factors might explain the lead time for an individual Jira issue, e.g., size, complexity, number of teams involved, ownership of the component, and other contextual factors. Therefore, the role of TD on a single Jira issue seems very far-fetched, although we plan to tackle this analysis in further work. So, considering the monthly average lead time value for Jira issues for each component helps mitigate the biases of such factors as mentioned above since it will allow us to observe trends on TDD and lead time. In addition, we want to investigate the impact of TDD on lead time from a bigger perspective, as both of them are context-dependent.

\begin{table}[htbp]
\captionsetup{labelsep=colon}
\caption{Description of the selected components.}
\label{components}
\centering 
\scriptsize 
\begin{tabular} {p{1.15cm} p{1.25cm} p{1.25cm} p{1.25cm} p{1.25cm} }
\toprule

Components & Individual Jira issues & Data points (Monthly) & Avg.monthly lead time (Days)  & Avg.monthly TDD (Minutes / KLOC) \\
\midrule
Pangolins & 78 & 23 & 9.18 & 187.45 \\
Mongoose & 98 & 22 & 3.41 & 96.77 \\
Zorilla & 211 & 24 & 4.14 & 59.45 \\
Shark & 194 & 22 & 3.82 & 57.27 \\
Sloth & 729 & 23 & 6.93 & 91.49 \\
Penguin & 157 & 23 & 7.73 & 269.49 \\
\bottomrule
\end{tabular}
\end{table}

We gathered all the Jira Issues for the years 2022 and 2023\footnote{We only consider this period because reliable lead time data became available at the company starting in early 2022, despite the components having a longer development history.}. 
In addition to this, we collected lead time data separately from the Jira endpoint for Jira issues. This data includes the issue key, creation time, and the time developers spend in various statutes like time in progress, open, review, testing, and more. However, based on the company's way of tracking lead time, as mentioned in Section~\ref{section:background_leadtime}, we calculate the lead time as an aggregated sum of time spent by issues in \emph{progress}, \emph{code review}, and \emph{testing}.





\subsection{Data Analysis}

We started by visualizing TDD to observe the nature of its distribution. Some of the Jira issues were linked with multiple components, so we removed those issues as we analyzed each component separately. In addition, a few of the lead time data points were notably high. The reason behind those long lead times could be low-priority issues, controversy, long code-review processes, pull-request re-scoping, and rejections. We discussed these extreme cases with the development team, and upon their recommendation after review, we treated them as outliers and removed the top five percentile of lead time data.

As shown in Figure~\ref{fig: final lead time}, lead time data was positively skewed to one-tail, we tried to transform the data to a more normally distributed form by applying square root transformation. We carried out a Shapiro–Wilk test \cite{gonzalez2019shapiro}, which is used for testing the normality of data distribution.

We performed different correlation tests such as Pearson, Spearman, and Kendal correlation coefficients, which were also adopted in a related study~\cite{lenarduzzi_technical_2021}. In addition, we plotted the TDD per thousand lines of code versus the monthly average lead time to resolve Jira issues for each component, which can be seen in Figure~\ref{fig_result_fix}, with each figure corresponding to a different component. Pearson correlation is a statistical measurement of the strength and direction of linear correlation between two variables~\cite{4459449}. Spearman and Kendall's rank correlation are nonparametric correlation metrics~\cite{croux_influence_2010} that measure the monotonic (but not necessarily linear) relationship between the two variables~\cite{puth2015effective}. Spearman correlation measures the degree of monotonic association between variables, and Kendall correlation compares the ranking of the values between the variables~\cite{lenarduzzi_technical_2021, puth2015effective}.

Correlation coefficient~\cite{ratner_correlation_2009} ranges in the interval $[-1, +1]$, where 0 indicates no linear relationship and $+1$ indicates a perfect positive linear relationship, which means that as one independent variable increases, the dependent variable also increases. On the other hand, $-1$ indicates a perfect negative linear relationship, meaning that the dependent variable decreases as the independent increases. Values between 0 and 0.3 (0 and $-0.3$) show a weak positive (negative) linear relationship. Likewise, values between 0.3 and 0.7 ($-0.3$ and $-0.7$) indicate a moderate positive (negative) linear relationship, while values between 0.7 and 1.0 ($-0.7$ and $-1.0$) suggest a strong positive (negative) linear relationship~\cite{ratner_correlation_2009}.

In order to answer our research question and test hypothesis, we then calculate the coefficient of determination, also called R$-$squared(R\textsuperscript{2})~\cite{ratner_correlation_2009}. The R$-$squared value indicates the percentage of variation in the dependent variable (lead time) explained by the independent variable (TDD). In addition, we also tried to fit different polynomial regression models with order 1,2,3 and calculated the RMSE (root mean squared value) for each order. The fitted models are visualized for each component and available in a GitHub repository~\footnote{\url{https://github.com/bhuwanpaudel/Impact-of-Technical-Debt-on-Lead-Time.git }\label{fn:footnote4}}. The smaller the RMSE value, the better the model~\cite{ratner_correlation_2009}.

\section{Results}\label{sec:results}

We began our analysis by gathering all Jira issues for 2022-2023. The total number of Jira issues over that period was \num{17544}. After removing Jira issues that were linked with multiple components, \num{15592} issues were left.
We started with ten components for this study; their total Jira issues were \num{1735}. Out of those components, we further analyzed only six that met our selection criteria (i.e., having more than twenty months of active development in the last two years). 

\begin{figure}[htbp]
  \centering
  \footnotesize 
    \begin{minipage}{0.95\columnwidth}
    \centering
      \includegraphics[width=\linewidth]{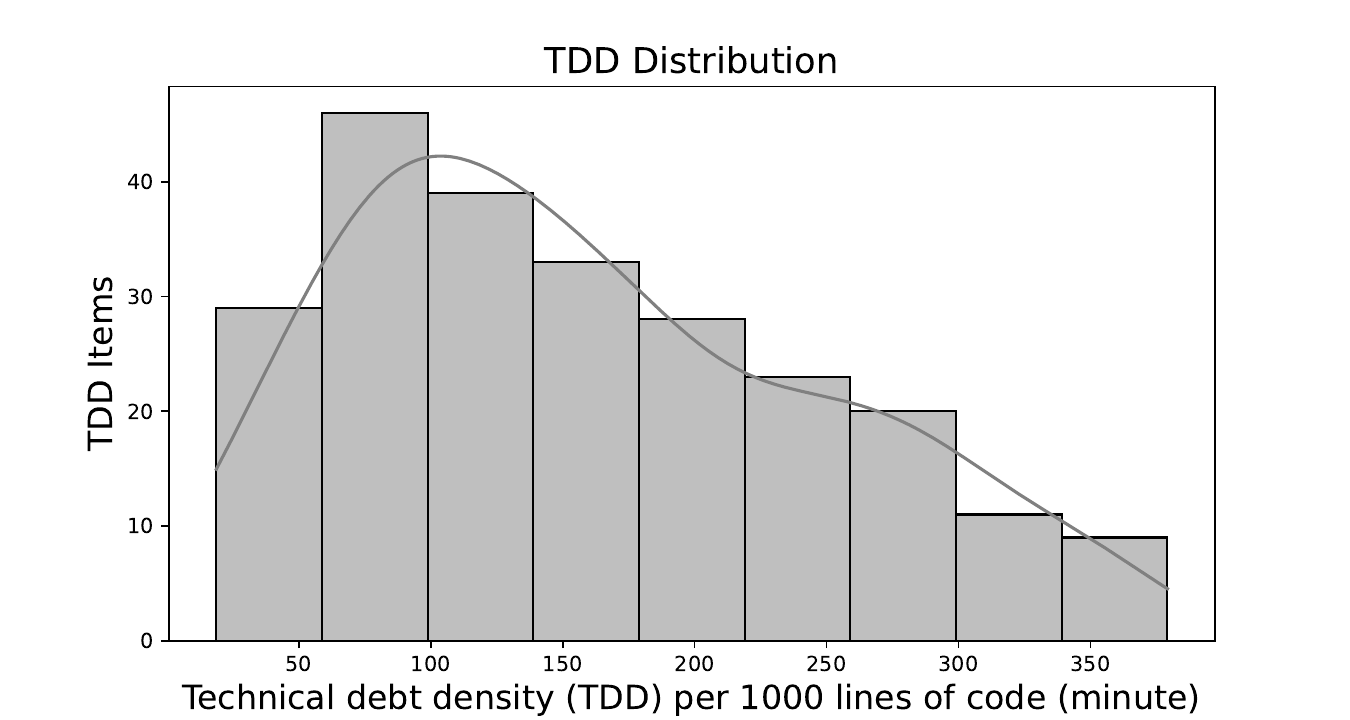}
      \caption{TDD distribution for selected components.}
      \label{fig: TD selected Repos}
    \end{minipage}

  \begin{minipage}{0.95\linewidth}
      \centering
      \includegraphics[width=\linewidth]{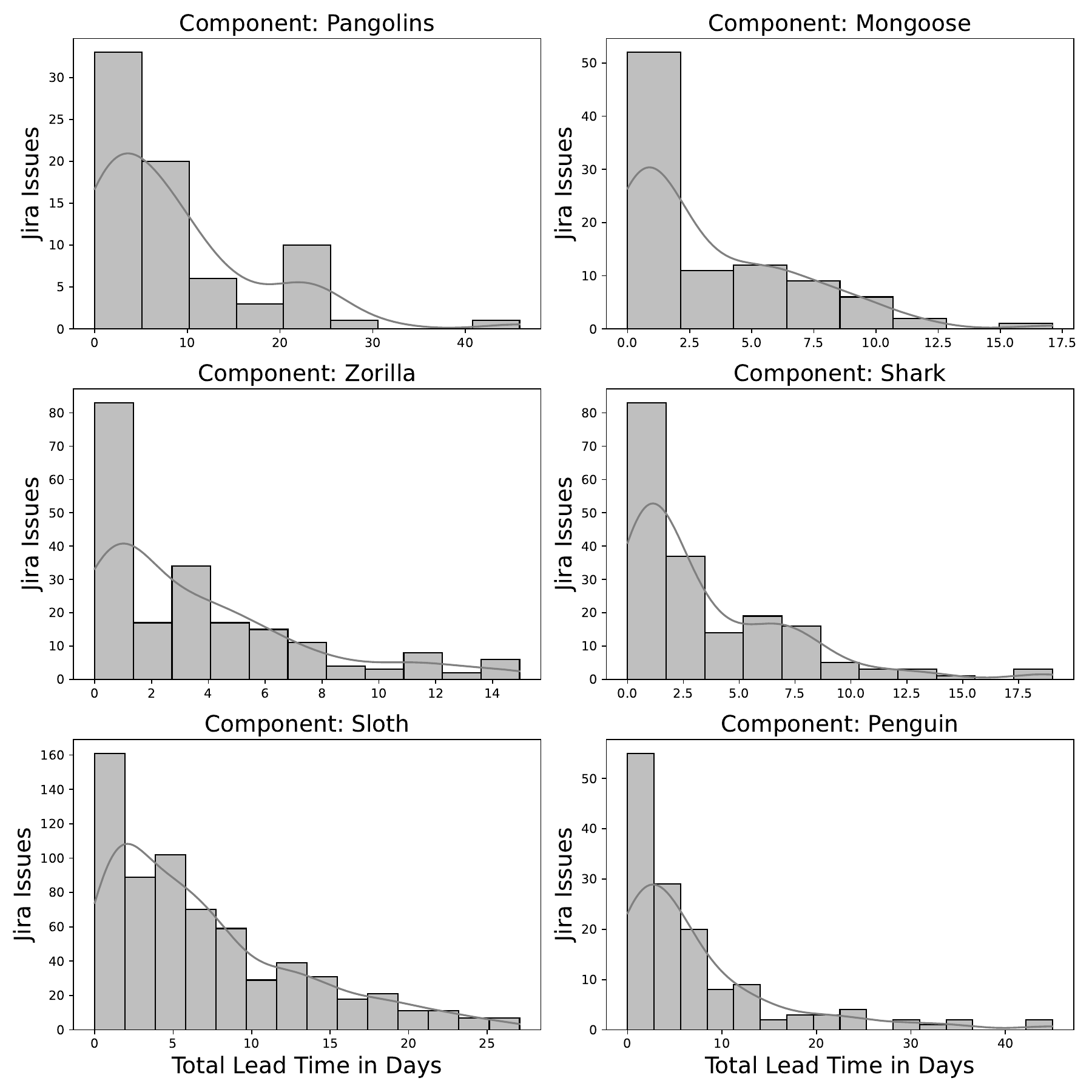}
    \caption{Lead time (days) distribution after removing top five percentile (outliers).}
    \label{fig: final lead time}
    \end{minipage}
\end{figure}

Figure~\ref{fig: TD selected Repos} shows that the TDD data nearly follows a normal distribution. However, lead time data had some outliers, so we removed the top five percentile of lead time data points. We then checked the distribution of lead time, which was not normally distributed, as shown in Figure~\ref{fig: final lead time}. We transformed it into a normally distributed form by applying a square root transformation. However, data still remained non-normally distributed, as tested by the Shapiro-Wilk test. The p-values for five of the components were significantly less than 0.05, leading to the rejection of the null hypothesis. Hence, we opted to work with non-transformed data to maintain the readability of results in the figures and the nature of data distribution. The results of the test are presented in Table~\ref{results}. 

\begin{table*}[htbp]
\captionsetup{labelsep=colon}
\caption{Results of the different tests performed in this study.}
\label{results}
\centering 
\scriptsize 
\begin{tabular}{p{1.40cm} | p{0.80cm} | p{1.10cm} | p{0.80cm} | p{0.70cm} | p{1.10cm} | p{0.7cm} | p{0.70cm} | p{1.10cm} | p{0.6cm} | p{1.6cm}}
\toprule
\multirow{2}{*}{\textbf{Components}} & \multicolumn{3}{c|}{\textbf{Correlations}} & \multicolumn{3}{c|}{\textbf{R-squared (R\textsuperscript{2})}} & \multicolumn{3}{c|}{\textbf{RMSE}} & \multirow{2}{1.60cm}{\textbf{Shapiro-wilk (P-value)}} \\
\cline{2-10}
\ & \textbf{Pearson} & \textbf{Spearman} & \textbf{Kendall} & \textbf{Linear} & \textbf{Quadratic} & \textbf{Cubic} & \textbf{Linear} & \textbf{Quadratic} & \textbf{Cubic} & \\
\hline

Pangolins & 0.01 & 0.08 & 0.05 & 0.00 & 0.07 & 0.11 & 6.29 & 6.04 & 5.94 & 0.09 \\
Mongoose & 0.28 & 0.18 & 0.13 & 0.08 & 0.29 & 0.41 & 2.29 & 2.01 & 1.83 & $1.01 \times 10^{-5}$ \\
Zorilla & 0.55 & 0.66 & 0.46 & 0.31 & 0.36 & 0.37 & 1.95 & 1.88 & 1.85 & $2.33 \times 10^{-5}$ \\
Sloth & $-0.41$ & $-0.28$ & $-0.21$ & 0.17 & 0.26 & 0.27 & 1.14 & 1.08 & 1.07 & $5.06 \times 10^{-6}$ \\
Shark & $-0.34$ & $-0.48$ & $-0.34$ & 0.12 & 0.12 & 0.13 & 1.80 & 1.80 & 1.78 & 0.0010 \\
Penguin & $-0.20$ & $-0.16$ & $-0.10$ & 0.04 & 0.05 & 0.05 & 5.81 & 5.78 & 5.78 & $3.72 \times 10^{-8}$ \\
\bottomrule
\end{tabular}
\end{table*}

\begin{figure*}[htbp]
    \centering
        \includegraphics[width=0.91\linewidth]{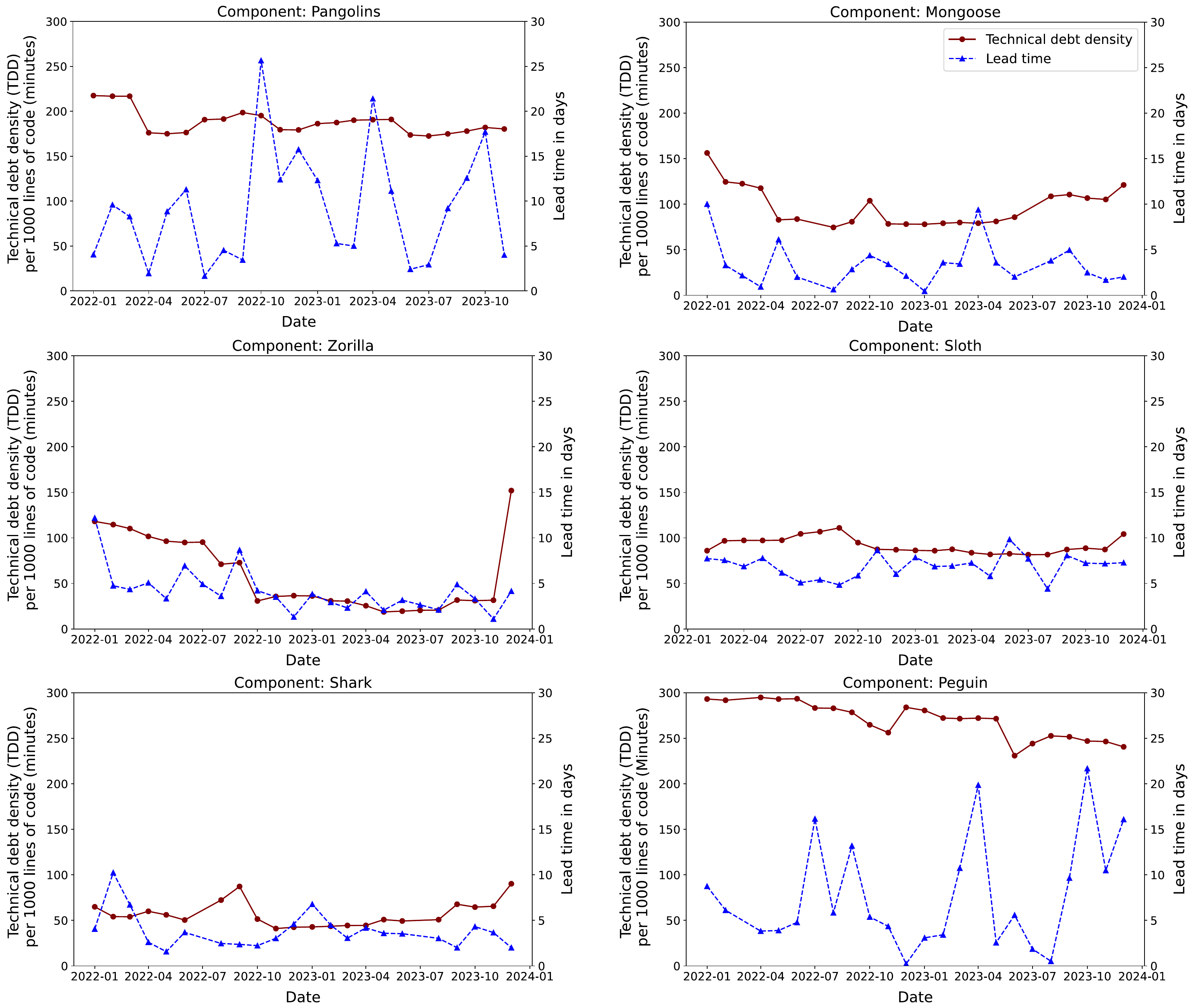}
    \caption{Technical debt density (TDD) per KLOC (minutes) vs average monthly lead time (days) for the components. Legend on the top-right graph is applicable to all.}
    \label{fig_result_fix}
\end{figure*}

Before calculating the correlation coefficients, we plotted each component's lead time and TDD data. Figure~\ref{fig_result_fix} illustrates the graphical representation of two variables over two years. 
After visualization, we calculated the different types of correlation coefficients, Pearson, Spearman, and Kendall, to test the correlation and association between the two variables for each component. Three of the six components had positive correlations, while the others had negative correlations. To be more precise, the component Pangolins showed a weak positive or almost no correlation. The components Mongoose showed almost a moderate positive correlation, while Zorilla showed a perfect moderate positive correlation. In contrast, component Penguin showed a weak negative correlation, while component Shark and Sloth showed a moderate negative correlation. The values of each correlation coefficient for all components are presented in Table~\ref{results}.

Once we calculated the correlation coefficients, we conducted the regression analysis to determine the percentage of variation in lead time that TDD can explain. To do this, we calculated the R-squared value, also known as the coefficient of determination for different orders. The R-squared value ranges from 0 to 1. A value of 0 means none of the variance is explainable, while a value of 1 means 100\% of the variance is explainable. The fitted lines for each order were then visualized and available in the GitHub repository\footref{fn:footnote4}. Moreover, we also calculated the root mean squared (RMSE) value to decide which order fits the most. 

Table~\ref{results} shows that the R-squared value for the component Pangolins was very low, while RMSE is high for both the linear and quadratic regression models. However, the R-squared value was 0.11 with 5.94 RMSE for the cubic regression. Similarly, for the components Mongoose and Zorilla, R-squared values for linear regression were 0.08 and 0.31, with RMSE values of 2.29 and 1.95, respectively. In contrast, cubic regression had values of 0.41 and 0.37, with RMSE values of 1.83 and 1.85, respectively. Despite negative correlations, the other three components (Sloth, Shark, and Penguin) had R-squared values of 0.17, 0.12, and 0.04 for the linear regression model, while the cubic regression values were 0.27, 0.13, and 0.05, respectively.


\section{Discussion}\label{sec:discussion}

In this section, we discuss our research contribution and results with regard to our research question and their implication for future research and practice.

We answered our RQ by investigating how TD impacts lead time to resolve Jira issues and what percentage of changes in lead time is explainable by TDD. To answer this question, we analyzed six individual components of a large-scale company separately.

We found that out of the six components studied, two components with higher technical debt density (i.e., low code quality) did not show a significant correlation between TDD and lead time. This aligns with the result of study~\cite{lenarduzzi_technical_2021}, which also did not find any meaningful correlation between TD violations and lead time. Additionally, our results indicate that the TDD in those two components explains 10\% or less of the lead time variance to resolve Jira issues. 

In the other two components with a low TDD (i.e., better code quality), we found a moderate positive correlation between TDD and lead time, meaning the lead time can be affected by the TDD. The linear regression analysis showed that the TDD can explain around 8\% and 31\% of the lead time changes. On the other hand, cubic regression analysis indicated that around 41\% and 37\%  of variance in lead time can be explained by TDD. The RMSE values for cubic regression were 1.83 and 1.85, respectively, much less than those for linear and quadratic regression. One of the studies by Besker et al.~\cite{besker2019software} also mentioned that around 23\% of developers' time is wasted due to TD. Likewise, a study by Tornhill \& Borg ~\cite{tornhill_code_2022} revealed that the time to resolve Jira issues in alert code quality is 124\% longer than in high code quality. Yet another study by Besker et al.~\cite{8094405} found that an average of 36\% of development time is wasted because of repaying TD interest.

Interestingly, we found a moderate negative correlation between the two variables in the two of the remaining components. However, the linear regression analysis showed that the TDD can only explain around 17\% and 12\% of the lead time changes. On the other hand, cubic regression analysis indicated that around 27\% and 13\%  of variance in lead time can be explained by TDD. 

Our findings suggest that the relationship between TDD and lead time in resolving Jira issues is either very weak or nonexistent if TD and lead time are aggregated monthly. TDD vs lead time is not always explainable and positive but rather unexplainable and negative in some components. The R-squared and RMSE values indicate that this relationship is more cubic than linear across all components studied, with positive and negative correlations, as shown in Table~\ref{results}. This indicates that the lead time to resolve Jira issues varies better in cubic than linear ways as TDD varies, meaning accelerated increases or decreases in lead time that are not proportionate.


In our study, we have found mixed results among components. Specifically, we did not see a clear relationship between TDD and lead time in components with higher TDD. In contrast, we saw a moderate positive/negative relationship in code with low TDD (i.e., high-quality code). So, the results suggest that it should be further explored in different contexts or across more components and data with unstudied variables that could affect the correlation between studied variables. Furthermore, exploring more ways to measure TD beyond static analysis tools like SonarQube would be better.

\section{Threats to validity}\label{sec:threats}

Our study, like other empirical studies, is subject to validity threats. This section discusses threats to validity from four aspects as defined by Runeson \& Höst~\cite{runeson_guidelines_2009}: construct validity, internal validity, external validity, and reliability. 

\textit{Construct validity} refers to the extent to which measures studied in the study represent what researchers intend to measure and what is being studied in accordance with the research question~\cite{runeson_guidelines_2009}. To measure the TDD, we relied on SonarQube, which is one of the most used and highly adopted tools to measure the TD~\cite{lenarduzzi_survey_2020, lenarduzzi_technical_2019} and has been used for some other related studies such as the study~\cite{lenarduzzi_technical_2021}. We have normalized the technical debt by the component size, eliminating the potential impacts of external factors such as size. Moreover, we have only considered the TD for the issues marked as open. This ensures that we obtain an accurate representation of the technical debt at the start of each month. 
The lead time for each issue is calculated based on the time developers spend on three statuses in Jira: time in progress, code review, and testing. This approach assures that we only considered the actual time developers spent resolving a Jira issue. We then calculated the average lead time for Jira issues in a month, eliminating the threat of known and unknown external factors affecting lead time, such as size, complexity, and ownership of the changed code. 

\textit{Internal validity} concerns the possibility of other factors affecting the main factors being studied. In this study, the major concern regarding internal validity is the correlation between the lead time and the TD. We have found a mixed correlation ranging from weak to moderate for different components. However, we acknowledge that several confounding variables we could not consider in this study might influence this correlation. We tried to mitigate this threat by normalizing TD and averaging lead time on a monthly basis. Nevertheless, further studies should be conducted to investigate the impact of such internal and external factors.

\textit{External validity} concerns the generalizability of findings. This case study is based on a large-scale financial product development company and may not represent the entire software development landscape. Even though we consider all the programming languages used, such as Klotin, TypeScript, Java, and JavaScript, Java was the main programming language. SonaQube provides a unique set of TD issues for each programming language. Moreover, we do not rely solely on statistical generalization when interpreting data. Rather, we analyzed and discussed each component separately. 

\textit{Reliability} is concerned with reducing the errors and biases in a study, ensuring data and analysis are independent of the specific researchers. In order to mitigate this threat, we had multiple discussions with our partner company regarding the data collection, analysis, and interpretation of the result. We reported our results using descriptive statistics and correlation models. Additionally, we also applied polynomial regression with order 1,2,3. We used standard Python packages to perform all statistical analyses. Furthermore, we engaged in pair programming and discussions to ensure the quality of our findings and increase our confidence in the results.

\section{Conclusion and Future Work}\label{sec:conclusions}

This study explores the relationship between TDD and lead time for resolving Jira issues and the extent to which lead time to resolve Jira issues is explainable by the TDD. Such an approach allows us to identify which components are most affected by TDD in terms of lead time, thus enabling better prioritization and informed decisions for software maintenance.

We analyzed six components to study the impact of TD (measured as TDD) on lead time in an industrial case. We obtained a set of mixed results. In two components, we found a moderate positive correlation with 37\% and 38\% of lead time variance (cubic) explained by TDD. We observed a negative moderate correlation in the other two components, while no meaningful correlation was seen in the remaining two.  

We got inconclusive results as it appears that TD is not the only factor affecting lead time. The reasons behind this might be more complex than it seems because the TD that exists now may impact lead time in the future. It means that TD might have a residual effect that could manifest later. Conversely, lead time could also be a source of TD or compromised code quality. For example, software developers might use sub-optimal solutions to be faster. Additionally, there could be other factors that influence lead time. However, exploring those factors does not seem feasible at the moment because the company started collecting data with traceability between Jira issues and code commits after we started our analysis.

Therefore, we plan to conduct further research to investigate other potential confounding factors that might affect this relationship, such as the number of affected lines of code for a Jira issue, who owns the code, or maybe the complexity of the change made. We are currently exploring such variables. We also plan to explore this relationship at a file level by including all the above-mentioned potential confounding variables.

\section*{Acknowledgement}
\noindent{This research was supported by the KKS Foundation through the KKS SERT Research Profile project (Ref. 2018010).}

\bibliographystyle{IEEEtran}
\bibliography{code_quality_vs_lead_time.bib}

\begin{thebibliography}{10}
\providecommand{\url}[1]{#1}
\csname url@samestyle\endcsname
\providecommand{\newblock}{\relax}
\providecommand{\bibinfo}[2]{#2}
\providecommand{\BIBentrySTDinterwordspacing}{\spaceskip=0pt\relax}
\providecommand{\BIBentryALTinterwordstretchfactor}{4}
\providecommand{\BIBentryALTinterwordspacing}{\spaceskip=\fontdimen2\font plus
\BIBentryALTinterwordstretchfactor\fontdimen3\font minus \fontdimen4\font\relax}
\providecommand{\BIBforeignlanguage}[2]{{%
\expandafter\ifx\csname l@#1\endcsname\relax
\typeout{** WARNING: IEEEtran.bst: No hyphenation pattern has been}%
\typeout{** loaded for the language `#1'. Using the pattern for}%
\typeout{** the default language instead.}%
\else
\language=\csname l@#1\endcsname
\fi
#2}}
\providecommand{\BIBdecl}{\relax}
\BIBdecl

\bibitem{besker2017time}
T.~Besker, A.~Martini, and J.~Bosch, ``Time to pay up: Technical debt from a software quality perspective.'' in \emph{CIbSE}, 2017, pp. 235--248.

\bibitem{lenarduzzi_technical_2021}
V.~Lenarduzzi, A.~Martini, N.~Saarimäki, and D.~A. Tamburri, ``Technical debt impacting lead-times: An exploratory study,'' in \emph{2021 47th euromicro conference on software engineering and advanced applications (seaa)}.\hskip 1em plus 0.5em minus 0.4em\relax {IEEE}, 2021, pp. 188--195.

\bibitem{zabardast_refactoring_2020}
E.~Zabardast, J.~Gonzalez-Huerta, and D.~Šmite, ``Refactoring, bug fixing, and new development effect on technical debt: An industrial case study,'' in \emph{2020 46th Euromicro Conference on Software Engineering and Advanced Applications ({SEAA})}.\hskip 1em plus 0.5em minus 0.4em\relax {IEEE}, 2020, pp. 376--384.

\bibitem{besker2019software}
T.~Besker, A.~Martini, and J.~Bosch, ``Software developer productivity loss due to technical debt—a replication and extension study examining developers’ development work,'' \emph{Journal of Systems and Software}, vol. 156, pp. 41--61, 2019.

\bibitem{tornhill_code_2022}
A.~Tornhill and M.~Borg, ``Code red: the business impact of code quality - a quantitative study of 39 proprietary production codebases,'' in \emph{Proceedings of the International Conference on Technical Debt}, ser. {TechDebt} '22.\hskip 1em plus 0.5em minus 0.4em\relax ACM, 2022-08-16, pp. 11--20.

\bibitem{8094405}
T.~Besker, A.~Martini, and J.~Bosch, ``The pricey bill of technical debt: When and by whom will it be paid?'' in \emph{2017 IEEE International Conference on Software Maintenance and Evolution (ICSME)}, Sep. 2017, pp. 13--23.

\bibitem{10449672}
P.~Avgeriou, I.~Ozkaya, A.~Chatzigeorgiou, M.~Ciolkowski, N.~A. Ernst, R.~J. Koontz, E.~Poort, and F.~Shull, ``Technical debt management: The road ahead for successful software delivery,'' in \emph{2023 IEEE/ACM International Conference on Software Engineering: Future of Software Engineering (ICSE-FoSE)}, 2023, pp. 15--30.

\bibitem{giardino_software_2015}
C.~Giardino, N.~Paternoster, M.~Unterkalmsteiner, T.~Gorschek, and P.~Abrahamsson, ``Software development in startup companies: the greenfield startup model,'' \emph{{IEEE} Transactions on Software Engineering}, vol.~42, no.~6, pp. 585--604, 2015, publisher: {IEEE}.

\bibitem{6224279}
F.~Khomh, T.~Dhaliwal, Y.~Zou, and B.~Adams, ``Do faster releases improve software quality? an empirical case study of mozilla firefox,'' in \emph{2012 9th IEEE Working Conference on Mining Software Repositories (MSR)}, 2012, pp. 179--188.

\bibitem{lehman1980programs}
M.~M. Lehman, ``Programs, life cycles, and laws of software evolution,'' \emph{Proceedings of the IEEE}, vol.~68, no.~9, pp. 1060--1076, 1980.

\bibitem{yu2013empirical}
L.~Yu and A.~Mishra, ``An empirical study of lehman’s law on software quality evolution,'' \emph{International Journal of Software and Informatics}, 2013.

\bibitem{9234106}
G.~Digkas, A.~Chatzigeorgiou, A.~Ampatzoglou, and P.~Avgeriou, ``Can clean new code reduce technical debt density?'' \emph{IEEE Transactions on Software Engineering}, vol.~48, no.~5, pp. 1705--1721, 2022.

\bibitem{digkas2017evolution}
G.~Digkas, M.~Lungu, A.~Chatzigeorgiou, and P.~Avgeriou, ``The evolution of technical debt in the apache ecosystem,'' in \emph{Software Architecture: 11th European Conference, ECSA 2017, Canterbury, UK, September 11-15, 2017, Proceedings 11}.\hskip 1em plus 0.5em minus 0.4em\relax Springer, 2017, pp. 51--66.

\bibitem{avgeriou_reducing_2015}
P.~Avgeriou, P.~Kruchten, R.~L. Nord, I.~Ozkaya, and C.~Seaman, ``Reducing friction in software development,'' \emph{IEEE Software}, vol.~33, no.~1, pp. 66--73, 2015, publisher: {IEEE}.

\bibitem{tom_exploration_2013}
\BIBentryALTinterwordspacing
E.~Tom, A.~Aurum, and R.~Vidgen, ``\BIBforeignlanguage{en}{An exploration of technical debt},'' \emph{\BIBforeignlanguage{en}{Journal of Systems and Software}}, vol.~86, no.~6, pp. 1498--1516, Jun. 2013. [Online]. Available: \url{https://doi.org/10.1016/j.jss.2012.12.052}
\BIBentrySTDinterwordspacing

\bibitem{10.1007/978-3-031-04115-0_1}
E.~Klotins and T.~Gorschek, ``Continuous software engineering in the wild,'' in \emph{Software Quality: The Next Big Thing in Software Engineering and Quality}.\hskip 1em plus 0.5em minus 0.4em\relax Springer, 2022, pp. 3--12.

\bibitem{10011486}
E.~Zabardast, J.~Gonzalez-Huerta, and F.~Palma, ``The impact of forced working-from-home on code technical debt: An industrial case study,'' in \emph{2022 48th Euromicro Conference on Software Engineering and Advanced Applications (SEAA)}, 2022, pp. 298--305.

\bibitem{al2019evolution}
M.~A. Al~Mamun, A.~Martini, M.~Staron, C.~Berger, and J.~Hansson, ``Evolution of technical debt: An exploratory study,'' in \emph{2019 Joint Conference of the International Workshop on Software Measurement and the International Conference on Software Process and Product Measurement, IWSM-Mensura 2019, Haarlem, The Netherlands, October 7-9, 2019}, vol. 2476.\hskip 1em plus 0.5em minus 0.4em\relax CEUR-WS, 2019, pp. 87--102.

\bibitem{ernst_measure_2015}
N.~A. Ernst, S.~Bellomo, I.~Ozkaya, R.~L. Nord, and I.~Gorton, ``Measure it? manage it? ignore it? software practitioners and technical debt,'' in \emph{Proceedings of the 2015 10th Joint Meeting on Foundations of Software Engineering}.\hskip 1em plus 0.5em minus 0.4em\relax ACM, 2015-08-30, pp. 50--60.

\bibitem{cunningham_wycash_1992}
W.~Cunningham, ``The {WyCash} portfolio management system,'' \emph{{ACM} Sigplan Oops Messenger}, vol.~4, no.~2, pp. 29--30, 1992, publisher: {ACM} New York, {NY}, {USA}.

\bibitem{besker_systematic_2016}
T.~Besker, A.~Martini, and J.~Bosch, ``A systematic literature review and a unified model of {ATD},'' in \emph{2016 42th Euromicro Conference on Software Engineering and Advanced Applications ({SEAA})}, 2016, pp. 189--197.

\bibitem{jaspan2023defining}
C.~Jaspan and C.~Green, ``Defining, measuring, and managing technical debt,'' \emph{IEEE Software}, vol.~40, no.~03, pp. 15--19, 2023.

\bibitem{lenarduzzi_survey_2020}
V.~Lenarduzzi, A.~Sillitti, and D.~Taibi, ``A survey on code analysis tools for software maintenance prediction,'' in \emph{Proceedings of 6th International Conference in Software Engineering for Defence Applications}, ser. Advances in Intelligent Systems and Computing.\hskip 1em plus 0.5em minus 0.4em\relax Springer, 2020, pp. 165--175.

\bibitem{lenarduzzi_technical_2019}
V.~Lenarduzzi, N.~Saarimäki, and D.~Taibi, ``The technical debt dataset,'' in \emph{Proceedings of the 15th International Conference on Predictive Models and Data Analytics in Software Engineering}.\hskip 1em plus 0.5em minus 0.4em\relax {ACM}, 2019, pp. 2--11.

\bibitem{baldassarre_diffuseness_2020}
M.~T. Baldassarre, V.~Lenarduzzi, S.~Romano, and N.~Saarimäki, ``On the diffuseness of technical debt items and accuracy of remediation time when using {SonarQube},'' \emph{Information and Software Technology}, vol. 128, p. 106377, 2020-12.

\bibitem{noauthor_developer_2018}
\BIBentryALTinterwordspacing
``\BIBforeignlanguage{en-SE}{The {Developer} {Coefficient}},'' Sep. 2018. [Online]. Available: \url{https://stripe.com/en-se/newsroom/stories/developer-coefficient}
\BIBentrySTDinterwordspacing

\bibitem{5680918}
H.~Liu, Z.~Ma, W.~Shao, and Z.~Niu, ``Schedule of bad smell detection and resolution: A new way to save effort,'' \emph{IEEE Transactions on Software Engineering}, vol.~38, no.~1, pp. 220--235, 2012.

\bibitem{10.1145/1985362.1985370}
Y.~Guo and C.~Seaman, ``A portfolio approach to technical debt management,'' in \emph{Proceedings of the 2nd Workshop on Managing Technical Debt}, ser. MTD '11.\hskip 1em plus 0.5em minus 0.4em\relax New York, NY, USA: Association for Computing Machinery, 2011, p. 31–34.

\bibitem{li_systematic_2015}
Z.~Li, P.~Avgeriou, and P.~Liang, ``A systematic mapping study on technical debt and its management,'' \emph{Journal of Systems and Software}, vol. 101, pp. 193--220, 2015.

\bibitem{borg_u_2023}
M.~Borg, A.~Tornhill, and E.~Mones, ``U owns the code that changes and how marginal owners resolve issues slower in low-quality source code,'' in \emph{Proceedings of the 27th International Conference on Evaluation and Assessment in Software Engineering}.\hskip 1em plus 0.5em minus 0.4em\relax {ACM}, 2023, pp. 368--377.

\bibitem{runeson_guidelines_2009}
P.~Runeson and M.~Höst, ``Guidelines for conducting and reporting case study research in software engineering,'' \emph{Empirical Software Engineering}, vol.~14, no.~2, pp. 131--164, Apr. 2009.

\bibitem{gonzalez2019shapiro}
E.~Gonz{\'a}lez-Estrada and W.~Cosmes, ``Shapiro--wilk test for skew normal distributions based on data transformations,'' \emph{Journal of Statistical Computation and Simulation}, vol.~89, no.~17, pp. 3258--3272, 2019.

\bibitem{4459449}
J.~Benesty, J.~Chen, and Y.~Huang, ``On the importance of the pearson correlation coefficient in noise reduction,'' \emph{IEEE Transactions on Audio, Speech, and Language Processing}, vol.~16, no.~4, pp. 757--765, 2008.

\bibitem{croux_influence_2010}
C.~Croux and C.~Dehon, ``Influence functions of the {Spearman} and {Kendall} correlation measures,'' \emph{Statistical Methods \& Applications}, vol.~19, no.~4, pp. 497--515, Nov. 2010.

\bibitem{puth2015effective}
M.-T. Puth, M.~Neuh{\"a}user, and G.~D. Ruxton, ``Effective use of spearman's and kendall's correlation coefficients for association between two measured traits,'' \emph{Animal Behaviour}, vol. 102, pp. 77--84, 2015.

\bibitem{ratner_correlation_2009}
B.~Ratner, ``The correlation coefficient: {Its} values range between +1/-1, or do they?'' \emph{Journal of Targeting, Measurement and Analysis for Marketing}, vol.~17, no.~2, pp. 139--142, Jun. 2009.

\end{thebibliography}

\end{document}